\documentclass[aps,onecolumn,showpacs,preprintnumbers,floats]{revtex4}
\usepackage{amsfonts}
\usepackage{amsmath}
\usepackage{graphicx}
\usepackage{dcolumn}
\usepackage{bm}
\usepackage{natbib}

\begin{document}

\title{Epitaxial TbMnO$_{3}$ thin films on SrTiO$_3$ substrates: A structural study}

\author{C.J.M. Daumont$^{1}$, D. Mannix$^{2}$, Sriram Venkatesan$^{1}$, G. Catalan$^{3}$, D. Rubi$^{1}$, B.J. Kooi$^{1}$, J.Th.M. De Hosson$^1$, B. Noheda$^{1}$}

\address{$^{1}$Zernike Institute for Advanced Materials, University of Groningen, 9747 AG Groningen, The Netherlands}
\address{$^{2}$Institut N$\acute{e}$el, CNRS-UJF, 25 Avenue des Martyrs, 38042 Grenoble Cedex 9, France}
\address{$^{3}$Earth Science Department, Cambridge University, Cambridge CB2 3EQ, United Kingdom}

\begin{abstract}

TbMnO$_{3}$ films have been grown under compressive strain on (001)-oriented SrTiO$_{3}$ crystals. They have an orthorhombic structure and display the (001) orientation. With increasing thickness, the structure evolves from a more symmetric (tetragonal) to a less symmetric (bulk-like orthorhombic) structure, while keeping constant the in-plane compression thereby leaving the out-of-plane lattice spacing unchanged. The domain microstructure of the films is also revealed, showing an increasing number of orthorhombic domains as the thickness is decreased: we directly observe ferroelastic domains as narrow as 4nm. The high density of domain walls may explain the induced ferromagnetism observed in the films, while both the decreased anisotropy and the small size of the domains could account for the absence of a ferroelectric spin spiral phase.
\end{abstract}

\maketitle

\section{Introduction}
Multiferroics are a class of materials exhibiting coexistence of two or more ferroic orders\cite{Hill,Fiebig,Wilma,TokuraScience,Tokura}: (Anti-)ferromagnetism, (anti-)ferroelectricity, ferroelasticity and/or ferrotoroidicity. Multiferroic materials have been increasingly studied in recent years due to the possibility of substantial coupling between the ferroic properties within a single phase. The most interesting case for applications occurs when the magnetic and electric degrees of freedom are coupled, giving rise to a large magnetoelectric (ME) effect. This can eventually lead to a range of novel devices that use the control of the spontaneous magnetization (polarization) of a material with an electric (magnetic) field. Unfortunately, multiferroics are very rare and efforts have been made to discover and synthesize new multiferroic and magnetoelectric materials, as well as to understand the underlying mechanisms.

Among the possible mechanisms producing multiferroicity and large
magnetoelectric coupling in single phase materials, the induction of
ferroelectricity in spiral spin systems has been intensively
studied\cite{Kenzelmann,Mostovoy}. The orthorhombic perovskite
TbMnO$_{3}$ is the best known example of such a
system\cite{TokuraScience,Tokura,Kajimoto} and has also been shown
to present giant ME effect\cite{Kajimoto}. Among the rare earth
manganites, TbMnO$_{3}$ presents an intermediate Mn-O-Mn bond angle
and lies, in the temperature versus ionic radius phase diagram, in
between two different magnetic phases, displaying a complex magnetic
behavior\cite{Goto}. TbMnO$_{3}$ shows antiferromagnetic ordering
below T$_{N}$$\sim$ 40K, where the Mn spins align in an
incommensurate sinusoidal structure. By decreasing the temperature
further, the propagation vector of the sinusoidal spin structure
decreases until it locks at T$_{lock}$$\sim$ 28K, where the magnetic
structure changes into an spiral
structure\cite{Tokura,Kajimoto,Goto,Kimura} that propagates along
the b-axis. Due to the Dzyaloshinsky-Moriya(DM)
interaction\cite{Dagotto,Vanderbilt}, a spontaneous polarization
along the c-axis and a strong ME effect arises below T$_{lock}$.

For practical devices, multiferroics are often preferred in thin
film form. Moreover, the strain induced by the mismatch between the
film and the substrate lattice parameters can lead to improving the
materials properties with respect to the bulk. Interestingly,
although TbMnO$_3$ is by now a very well known material, only a few
reports on this perovskite in thin film form are
available\cite{Cui,Rubi,Kirby}. Since the magnetic and ferroelectric phases are
determined by the crystal structure, a thorough characterization of
the structure of the films is needed. This, however, is not straight
forward due to the small thickness of the films and grazing incidence
diffraction, using synchrotron sources, is required. In this work,
we present the growth and structural characterization of TbMnO$_3$
perovskite thin films with thicknesses lower than 100nm on single
crystals of (001)-SrTiO$_{3}$. (001)-SrTiO$_{3}$ is preferred in
studies of epitaxy in perovskites because it can be obtained with
atomically flat surfaces, favoring high quality growth. The films
are shown to be very flat, c-oriented and epitaxially strained.
Their unit cell is orthorhombic with an orthorhombic distortion that
can be tuned with the film thickness up to a thickness of about
50nm.

Although ferroelectric and magnetoelectric measurements are still in
progress, the magnetic properties (reported elsewhere\cite{Rubi})
suggest that these films are remarkably different from the bulk
material: No evidence of the spiral spin structure that gives rise
to multiferroic behavior was found in the magnetic data. On the
other hand, ferromagnetic interactions are induced in the films,
which renders them appealing for applications such as spin valves,
where ferromagnetic insulators are required. This is in agreement
with  a very recent work on the same system\cite{Kirby} and reports in other orthorhombic manganite films\cite{Marti}. We show here that
 the structure of the films and microstructure are likely to explain the
difference between the films and bulk behavior.

\begin{figure}[htb]
\centering
\includegraphics[width=12cm]{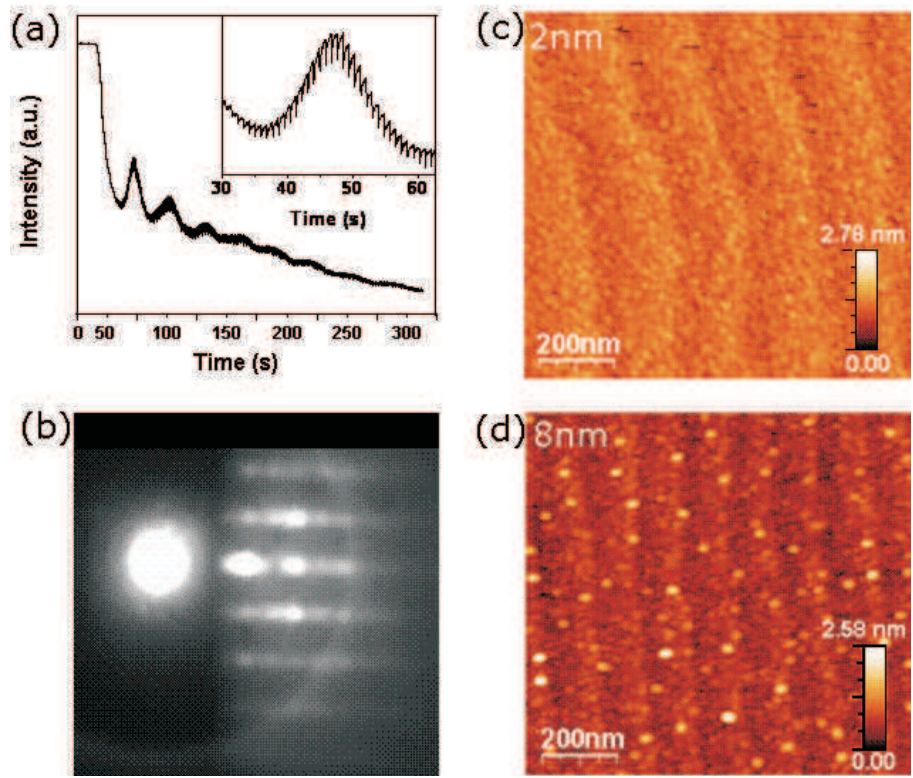}\\
\caption{(Color Online): (a)Intensity of the
specular spot of the RHEED pattern as a function of deposition time. The inset
is a blow-up of the first RHEED oscillation showing the laser pulses. (b) Typical RHEED
pattern obtained after growth and annealing. The picture is taken
after cooling down to room temperature and under vacuum. (c) and (d) AFM images of an 2nm
and 8nm thin film, respectively, grown at pO$_2$= 0.25 mbar. The vertical
and horizontal scales are shown at the bottom of the pictures}
\label{Figure2_0001}
\end{figure}

\section{Experimental}

(001)-oriented TbMnO$_{3}$ (TMO) thin films were deposited on
(001)-SrTiO$_{3}$ (STO) substrates by Pulsed Laser Deposition (PLD)
assisted by reflective high energy electron diffraction (RHEED),
using a KrF excimer laser with $\lambda=248nm$. A stoichiometric
target of TbMnO$_{3}$ was prepared by means of a standard
solid-state reaction. Prior to deposition, the substrates were
chemically treated and fired in O$_{2}$ in order to obtain TiO$_{2}$
single-terminated flat terraces\cite{Koster}. The deposition of the
films reported here was performed at 750$^{o}$C in oxygen pressures
of 0.25 mbar to 0.9 mbar. The laser fluence was 2 J/cm$^{2}$, with a
repetition rate of 1Hz, and a distance between substrate and target
of 55mm. After deposition, the films were slowly cooled down
(-3$^{o}$C/min) to room temperature under an oxygen pressure of 100
mbar.

The highly distorted perovskite TMO crystallizes in a orthorhombic structure (space group: Pbnm) with lattice constants of a$_o$=5.2931
{\AA}, b$_o$=5.8384 {\AA} and c$_o$=7.4025 {\AA} \cite{Alo00}. However, lattice parameters reported so far for the bulk/single crystal
system, and in particular the c-lattice parameter, vary significantly and this has been attributed to different growth conditions \cite{Kenzelmann,Blasco,Aliouane}. The STO substrate is a cubic perovskite with lattice constant of a=3.905 {\AA}. During growth, the intensities of the RHEED patterns change as a result of the relative surface coverage, roughening and disorder of the growing layers\cite{Ichimiya}. The intensity of the specular spot of the RHEED pattern was recorded from the beginning of the growth (see Figure 1(a)). With the first laser pulses, a large drop of the RHEED intensity takes place, indicating an important roughening of the surface with respect to the STO substrate. The recovery of intensity observed after approximately 35 laser pulses corresponds to the growth of the first complete layer of TMO.

The crystallinity and structure of the films were studied by standard x-ray diffraction (XRD) with a Panalytical X'Pert MRD diffractometer, while their surface morphology was probed by means of a Nanoscope IIIa atomic force microscope (AFM). X-ray photoelectron spectroscopy (XPS) experiments were performed in a SSX-100 spectrometer from Surface Science Instruments with a monochromatic Al-K x-ray source (h$\nu$=1486.6eV). XPS in films of different thickness and grown at different oxygen pressures are consistent with a manganese valence of +3 and no evidence of Mn4+ or Mn2+ was found \cite{Rubi}. In order to probe the in-plane lattice, grazing incidence x-ray diffraction (GIXD) was performed with synchrotron radiation, both at the W1 beamline in HASYLAB-DESY (Hamburg) and the XMaS beamline in ESRF (Grenoble). TEM plan-view specimens were prepared by conventional procedure involving cutting, grinding, polishing, dimpling and ion milling. A precision ion polishing system (Gatan model 491) with 4 Kv Ar+ beams was used, in which both guns make an angle of 8$^{o}$ with the bottom side of the substrate. On the other side, the film is covered with a piece of glass to avoid any contamination by redeposition from the sputtered backside. The observations have been performed with a JEOL 2010F electron microscope at an accelerating voltage of 200 kV.

\section{Results}

About eight RHEED intensity oscillations can be seen at the beginning of the growth corresponding to an initial layer-by-layer (2D-like) growth. However, those oscillations are superimposed on a decreasing intensity background, indicating that an overall roughening occurs as the growth proceeds. After the oscillations have faded away and the intensity of the specular spot has decreased to a certain intensity, the later remains constant during the rest of the deposition (outside of the graph range) and the pattern changes from a purely stripy pattern to the mixed
stripe-spot pattern shown in Figure 1(b). This type of behavior has been reported recently for other oxides as a 'pseudo-2D islands growth'\cite{Shin,Daudin}. This change from 2D character to 'pseudo'-3D character of the growth has also been identified using AFM. Indeed, Figure 1(c) shows the morphology of a 2nm film, for which the deposition was stopped shortly before the 2D-to-3D transition. It can be seen that the film is atomically flat, showing the step-and-terrace morphology of the substrate. However, when the deposition is stopped at the initial stage of the transition from 2D to 3D growth, particles with a diameter of around 30 nm start appearing, as shown in Figure 1(d). We found that the change from 2D to 3D growth occurs at a thickness in between 2 and 5 nm.

\begin{figure}[htb]
\centering
\includegraphics[width=12cm]{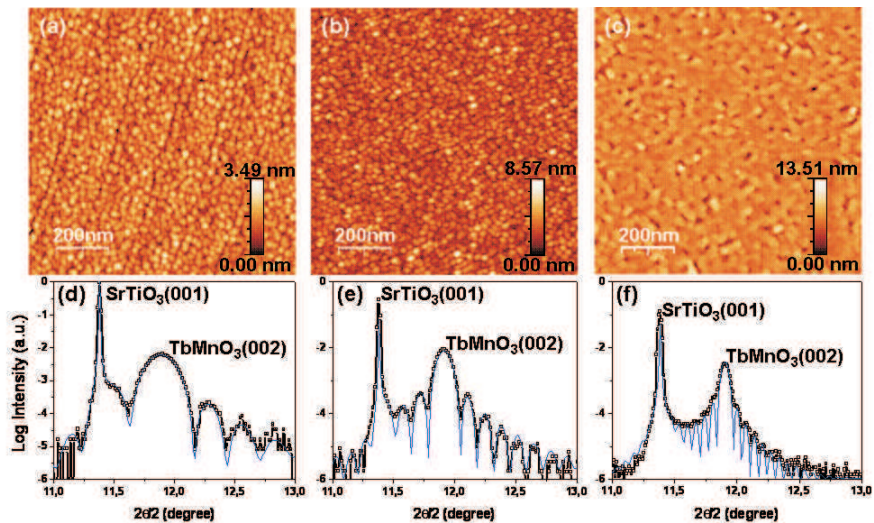}\\
\caption{(Color Online): AFM images of 17nm (a), 34nm (b) and 67nm (c) thick TMO thin films grown with a pO$_2$= 0.25 mbar. The scans area is 1$\mu $mx1$\mu $m. The corresponding diffraction patterns can be seen in (d), (e) and (f), respectively. The experimental data are are shown as thick black lines. The fit to the data is shown in each case as a thin blue line.} \label{Figure3_0001.eps}
\end{figure}

Figures 2(a) to 2(c) show AFM images of three TMO films with different thicknesses grown under the same conditions. In the 17nm film (Figure 2(a)), although it is already in the 3D growth regime and the grains dominate the film morphology, the steps from the substrate are still visible. The steps are not clearly visible for a 34nm film and a 67 nm (Figures 2(b),(c)). Figures 2(d)-(f) show typical 2$\theta$-$\omega$ x-ray diffractograms around the (001)$_c$ Bragg reflection of STO (the most intense ones in the pattern), for the same three films. The film peaks are those at the right hand-side of the STO reflections. The blue lines are the fits of the crystal truncation rod using a kinematical model. These calculations take into account, not only the complex refractive index of the substrate and the film, but also the angular dependent atomic scattering factors. The high quality and flatness of the interfaces is evidenced by the Laue fringes around the film peak for thicknesses up to 67nm.

Figure 3 shows a broader 2$\theta$-$\omega$ scan including both the
(001)$_c$ and the (002)$_c$ STO Bragg reflections. No secondary or
impurity phases could be observed. A phi-scan around the
(024)$_{pc}$ (ie (2$\overline{2}$8)$_o$ ) of TMO (see inset) shows
the peaks separated 90 degrees from each other, confirming the
four-fold symmetry of the film and the coherent growth. From the
position of the film peaks of figure 3, an out-of-plane lattice
spacing of 3.737 {\AA} was found. X-ray diffraction, thus, shows
that the films are single phase and (001)-oriented (with the
$c$-axis perpendicular to the surface). Using a value of Poisson
ratio of 0.35, typical in manganites\cite{Pois}, the out-of-plane
lattice spacing is estimated to expand from 3.70 {\AA} (c/2 of bulk
TMO) to about 3.74 \AA, in good agreement with our measurements.

\begin{figure}[htb]
\centering
\includegraphics[width=12cm]{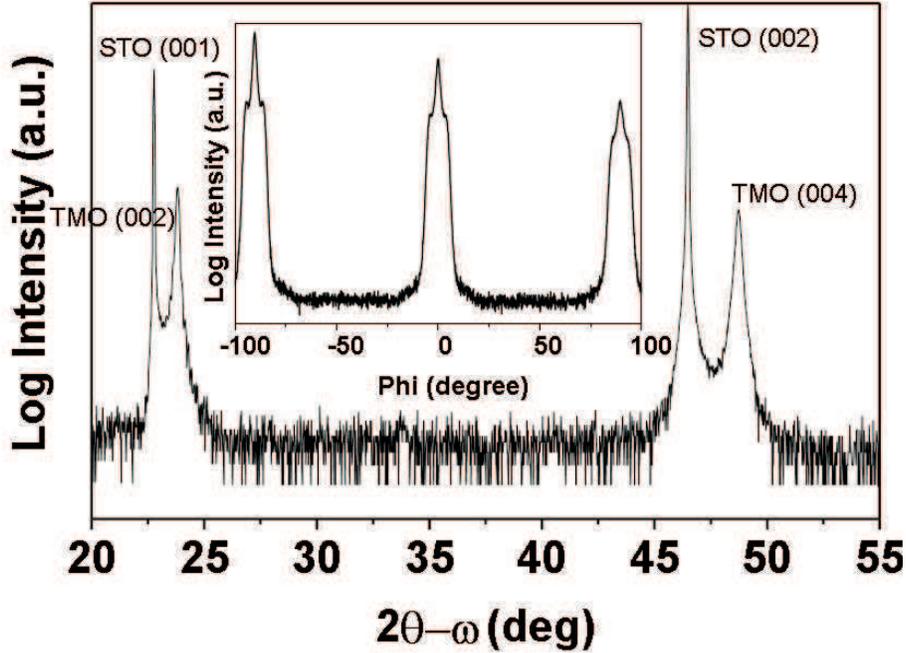}\\
\caption{2$\theta$-$\omega$ for a 67nm thick film of
TMO grown on STO. Inset: Phi-scan around the (2$\overline{2}$8)$_o$ reflection.} \label{Figure4_0001}
\end{figure}

A larger out-of-plane lattice parameter than the bulk's one could also be explained by a film whose strain is relaxed by means of oxygen vacancies, as they are known to expand the unit cell\cite{Babei,Rudman}. To investigate this possibility, the oxygen pressure during deposition was increased up to 0.9 mbar (the maximum attained in our set-up). Within the range of oxygen pressure investigated, the out-of-plane lattice parameter indeed decreases linearly with increasing oxygen pressure during deposition, consistent with the number of oxygen vacancies decreasing for increasing pressures. The maximum pressure of 0.9 mbar seems to be large enough to produce stoichiometric films since the unit cell of the relaxed layers is as large as the bulk's one, as it will be discussed later.

\begin{figure}[htb]
\centering
\includegraphics[width=12cm]{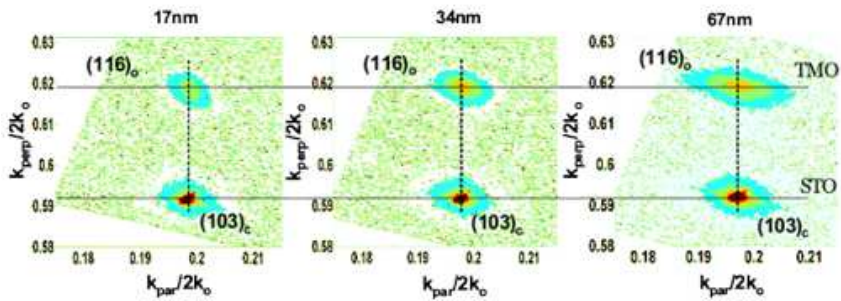}\\
\caption{(Color Online): Reciprocal space maps around the (103)$_c$
Bragg reflection of the STO substrate for 17nm, 34nm and 67nm films
grown at 0.25 mbar. The vertical and horizontal lines are guides to
the eyes. The abscissa (ordinate) represents the in-plane
(out-of-plane) component of the scattering vector. Both are
normalized by 2k$_o$=4$\pi$/$\lambda$.}
\label{Figure6_0001}
\end{figure}

A detailed investigation of the film structure was performed using x-ray diffraction to map selected areas of the reciprocal space. Figures 4(a)-(c) show reciprocal space maps (RSMs) around the (103)$_c$ Bragg reflections of a 17nm, 34nm and 67nm
thick film, respectively, grown at 0.25mbar. The peaks corresponding to the films (substrates) are those at larger (smaller) K$_{perp}$ (see horizontal lines in the figure). The in-plane and out-of plane components of the (103)$_{c}$ and (013)$_{c}$ TMO peaks do not change as a function of thickness and the in-plane lattice parameter remains identical to that of STO for thicknesses up to 67nm, indicating that the films are coherent with the substrate along the pseudo-cubic [100] or [010] directions. As the thickness is increased, only a broadening of the rocking curve of the TMO  film can be seen, indicating increase of mosaicity. However, when the RSM is taken around the (113)$_{c}$, a different picture is observed, as shown in figure 5(a) for a 67nm thin film. Instead of a coherent film peak, two film peaks are found that could be indexed as orthorhombic (206)$_o$ and (026)$_o$. The data therefore, fit a structural model in which the films are orthorhombically distorted, similar to the bulk material, but keep coherence with the cubic substrate along the [100] or [010] directions.

\begin{figure}[htb]
\centering
\includegraphics[width=12cm]{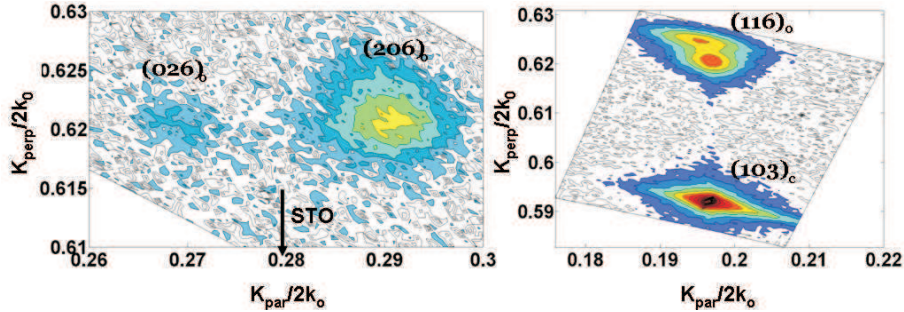}\\
\caption{(Color Online): (a)RSM around (206)$_o$ and (026)$_o$ for a 67nm thick TMO film grown at 0.25 mbar. The arrow indicates the k$_{par}$ of the STO substrate. (b)RSM around
the (103)$_c$ STO Bragg reflection for a 80nm thick TMO film grown
at 0.9 mbar. The abscissa (ordinate) represents the in-plane
(out-of-plane) component of the scattering vector, both normalized
by 2k$_o$=4$\pi$/$\lambda$.} \label{Figure7_0001}
\end{figure}

The orthorhombic ${a_o}$, ${b_o}$ and ${c_o}$ lattice parameters can
be extracted from the ${(116)_o}$ and ${(206)_o}$ reflections shown
above. Figure 6 shows the evolution of the lattice parameters as a
function of thickness for the films grown with pO$_2$= 0.9 mbar and
pO$_2$= 0.25 mbar. Both series display similar trend, showing that
the oxygen vacancies, which are likely to be present in the films
grown at the lowest pressure, giving rise to larger ${c_o}$ values,
do not change the general behavior. The orthorhombic distortion is
the smallest for the thinnest films and a 2nm thick TMO film shows a
tetragonal symmetry as evidenced by GIXD (not shown here). With
increasing thickness, ${c_o}$ remains unchanged, while ${a_o}$
decreases and ${b_o}$ increases towards their bulk value. However,
the bulk lattice parameters are not reached in a continuous way and
${a_o}$ and ${b_o}$ saturate for a thickness of around 50 nm. Above
about 70nm, a new spot appears in the diffraction maps, as shown in
Figure 5(b). This second spot corresponds to a relaxed unit cell
with bulk lattice parameters (dashed lines in Figure 6). The fact
that, in the films grown at 0.9mbar, this spot appears with lattice
parameters almost identical to the bulk ones, strongly indicates
that these films are stoichiometric, as discussed above. We have
shown that the films are orthorhombic but keep coherence with the
cubic substrate along the [010]$_c$ or [100]$_c$ directions. Thus,
the films orthorhombic lattice parameters fulfill
${a_{o}^{2}+b_{o}^{2}=(2a_{STO})^{2}}$ for all thickness and only
the in-plane angle $\gamma$ (see figure 9) changes with increasing
thickness. Therefore, the strain state is constant and so is the $c$
lattice parameter.

\begin{figure}[htb]
\centering
\includegraphics[width=8cm]{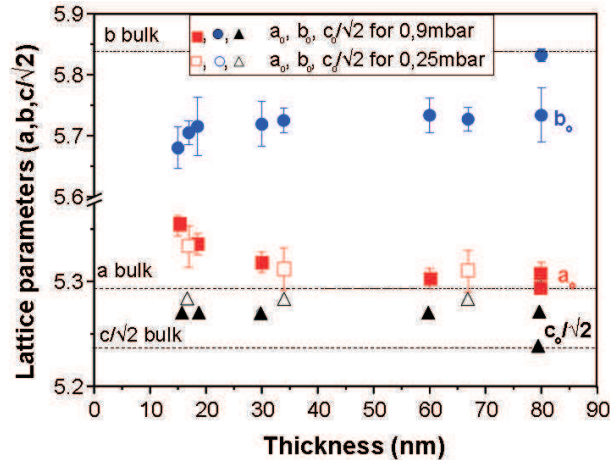}\\
\caption{Orthorhombic lattice parameters
(${a_o}$, ${b_o}$, ${c_o}$/$\sqrt{2}$) as a function of thickness for films grown with pO$_2$=
0.9 mbar (filled symbols) and 0.25mbar (open symbols). In the 80nm film, two phases are found and one of them (symbols marked with a star) corresponds to the bulk orthorhombic unit cell\cite{Alo00}. The lines are guides to the eye.} \label{Figure8_0001}
\end{figure}

In order to better understand the in-plane coherence of the films with the substrate and the domain structure of the film, GIXD experiments were performed using synchrotron radiation. Figure 7 (a)-(d) shows in-plane RSM's around selected reciprocal lattice points of the STO lattice for a 8 nm thin film grown at 0.25mbar. In all the maps, the central sharp spot corresponds to the STO Bragg peak and the other broader visible reflections around it are from the TMO film. As shown in (a), four broad peaks can be seen around the (110) substrate spot. They correspond to the different equivalent orientations of the orthorhombic unit cell (see sketch in figure 9). The presence of these four domains maintains in the film the four-fold symmetry of the cubic substrate. Moreover, as shown in figure 7(b) and (c), a peak broadening in the perpendicular direction (along the [010] for the (100) reflection and along [100] for the (010) reflection), the so-called rocking curve, is seen around the central cubic position. This corresponds to the part of the film that is coherent with the substrate (with lattice spacing 3.90 \AA). As seen in Figure 7(d), around the (220)$_c$, only two of the four peaks can be observed, those corresponding to the (400)$_o$ reflections. The absence of the (040)$_o$ peaks is due to the relative intensities of those reflections. Indeed, the intensity ratio I(200)$_o$/I(020)$_o$ is 1.4 in bulk, whereas a ratio of 77 is found for I(400)$_o$/I(040)$_o$. This makes the intensity of the (040)$_o$ too weak for us to detect it.

\begin{figure}[htb]
\centering
\includegraphics[width=8cm]{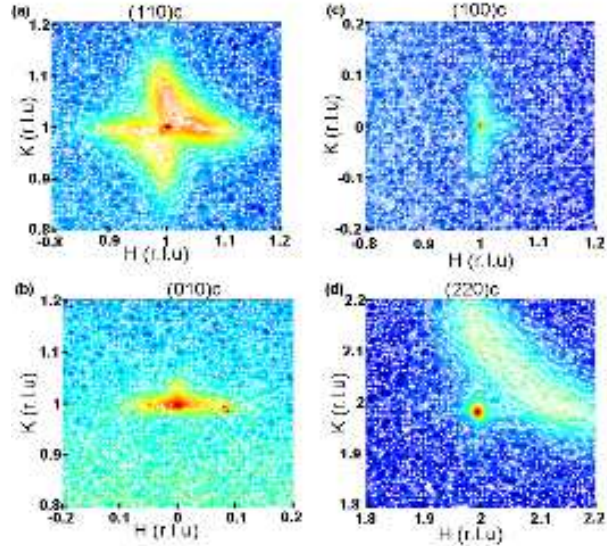}\\
\caption{(Color online): RSMs of a 8nm TMO film grown at 0.25mbar around: (a) the
(110)$_c$; (b) the (010)$_c$; (c) the (100)$_c$ and (d) the (220)$_c$ STO Bragg reflections. The axes are in reciprocal lattice units of the substrate (1 r.l.u.= 2$\pi$/3.905 \AA)}\label{Figure9_0001}
\end{figure}

\begin{figure}[htb]
\centering
\includegraphics[width=8cm]{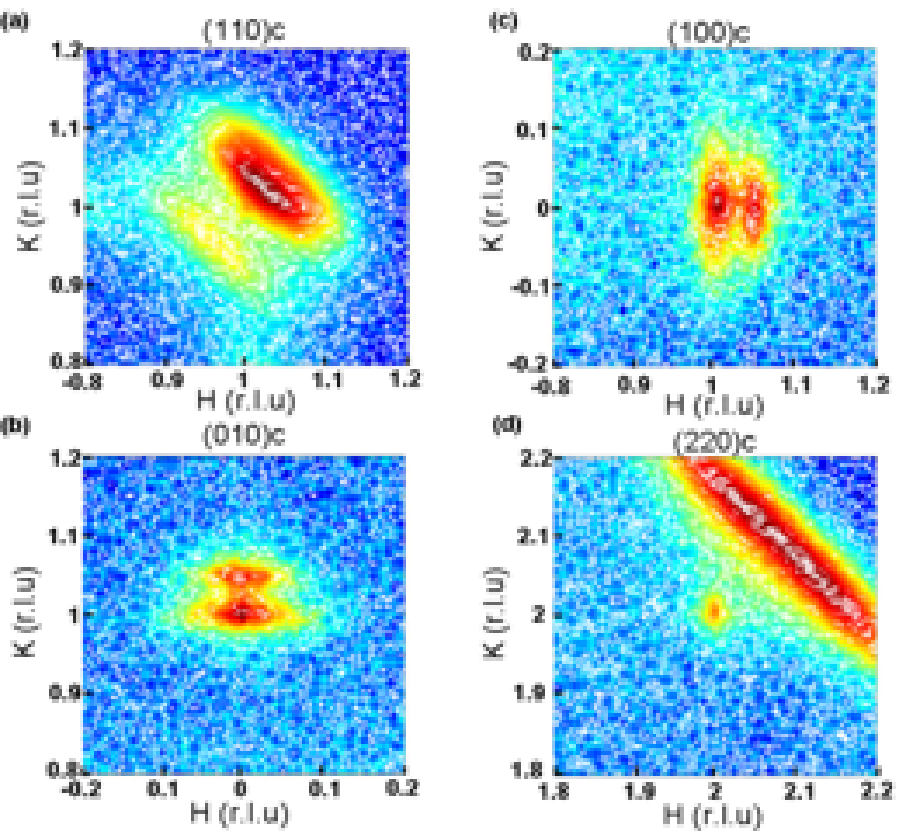}\\
\caption{(Color online): RSMs for a 55nm TMO film grown at 0.25mbar around (a) the
(110)$_c$; (b) the (010)$_c$; (c) the (100)$_c$ and (d) the (220)$_c$ STO Bragg reflections. The axes are in reciprocal lattice units of the substrate (1 r.l.u.= 2$\pi$/3.905 \AA)} \label{Figure8}
\end{figure}

GIXD experiments were also performed for a 55nm film grown at 0.25mbar. Figure 8 (a)-(d) shows different in-plane RSMs for this thicker film. From (a), only two film reflections are observed around the substrate (110) peak (which in this case is not visible due to the grazing incidence geometry and the larger thickness of the film). The peaks correspond to the (200)$_o$ (higher H and K values) and the (020)$_o$ (at lower H and K values). Analysis of these diffraction maps indicates that, along with the change of a$_o$ and b$_o$, a gradual rotation of the unit cell also occurs at larger thickness, so that the orthorhombic axes (a$_o$ and b$_o$) align parallel to the [110] of STO (see figure 9c). This can be better seen in figures 8 (a) and (d), where the extremely broad peak is due to domains existing at all orientations in between those sketched in figure 9(a) and (b) and those in (c). The coherence with the substrate in the thinner films is then given by [110]$_o$ parallel to [100]$_c$ (or to [010]$_c$). By further increasing the thickness, the strain energy has increased such that the coherency along the [100]$_c$ cannot be kept and the films relax to their bulk structure, with the orthorhombic in-plane axis parallel to the [110] of STO.

\begin{figure}[htb]
\centering
\includegraphics[width=8cm]{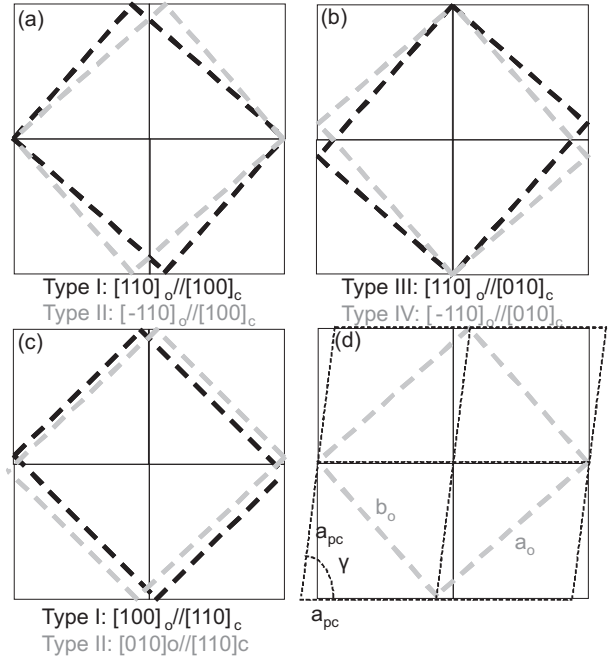}\\
\caption{(a) and (b)Sketch of the four types
of orthorhombic domains present in the thinner films and their coherence with the substrate.(c) Sketch of the orthorhombic domains present in the thicker films and their coherence with the substrate. d) Relationship between the orthorhombic and pseudo-cubic lattice parameters with the substrate lattice}\label{Figure9}
\end{figure}

Figure 10 shows TEM plan view images and electron diffraction patterns recorded for two samples with thicknesses of 17nm and 67nm. In the plan view image of the 17nm TMO film, two differently oriented types of rectangular domains can be seen along with strain contrast for domains having the same orientation. This is consistent with the four types of domains seen via XRD. For the thicker film, two orientations can still be seen along with bigger domains, consistent with the coalescence of the grains seen by AFM. Moreover, strain contrast is present within the domains showing that the domains coalesce as the thickness increases. Screw dislocations can be seen on the thicker film. In addition of screw dislocations at the interface, another type of defects probably exist at the domain boundaries that take care of the in-plane rotation of the domains. The electron diffraction patterns confirm what has been seen in x-ray diffraction: for the thinner film, the diffraction peaks are the superposition of four TMO domains along with the contribution from the substrate; for the 67nm TMO thin film, the electron diffraction pattern differs from that of the very thin films and shows a rotation gradient, as discussed above.

\section{Discussion}

For the 17nm film, the domains are found to have a small width of
about 4nm (see Fig. 10a). This domain size is particularly small
when compared with the typical thickness of magnetic domain walls,
which are of the order of several nm\cite{Hubert,Catalan}. This
means that in the thin films the fraction occupied by domain walls
is comparable to that of the domains themselves, and therefore the
properties of the domain walls are likely to affect the overall
magnetic properties, that is, the induced weak ferromagnetism, and
perhaps even the absence of a lock-in transition\cite{Rubi}. For
example, while the orthorhombic domains are known to be
antiferromagnetic in character, in agreement with the negative
Curie-Weiss temperature measured in the films\cite{Rubi}, the
magnetic interactions at the domain walls can give rise to a
ferromagnetic component. Symmetry arguments show that
magnetoelectric coupling can induce ferromagnetism in the domain
walls of ferroelectric antiferromagnets\cite{Privratska}, and works
by Fiebig and co-workers have also shown that the ferroelectric
domain walls of multiferroic hexagonal manganites can have a net
magnetization at their center\cite{Fiebig1} as well as enhanced
magnetoelectric coupling\cite{Fiebig2}.

The small domain size is likely to also affect the spin spiral. In
bulk TbMnO$_{3}$, the period of the spiral is of the order of 2nm, which
is comparable to the domain size and so it may well be destroyed by
the proximity of the domain boundaries. Moreover, even if the
spiral spin order that gives rise to the lock-in and ferroelectric
transition survived, the domain size may be inconmensurate with the
spiral period, meaning that some spins will not be fully
compensated, an effect that will be more noticeable for smaller
domains. The presence of small domains would also prevent long range
coherence, since the spiral propagation direction, which is along
the orthorhombic b-axis, has to to flick from x to y directions at
each domain wall.

Finally, the reduction of in-plane anisotropy means that the
difference between the a and b lattice parameters in the
orthorhombic structure decreases. For the extreme case of the
tetragonal films (2nm thick or less) there is no privileged
direction, and hence there can be no spin spiral. But even for the
thicker orthorhombic films, strain is likely to affect, or even
destroy, the spin spiral, since the changes in in-plane anisotropy
with respect to the bulk compound must also affect the Mn-O-Mn bond
angle, which is known to be the critical parameter for the
appearance of the lock-in spiral phase\cite{Kimura2}

\begin{figure}[htb]
\centering
\includegraphics[width=12cm]{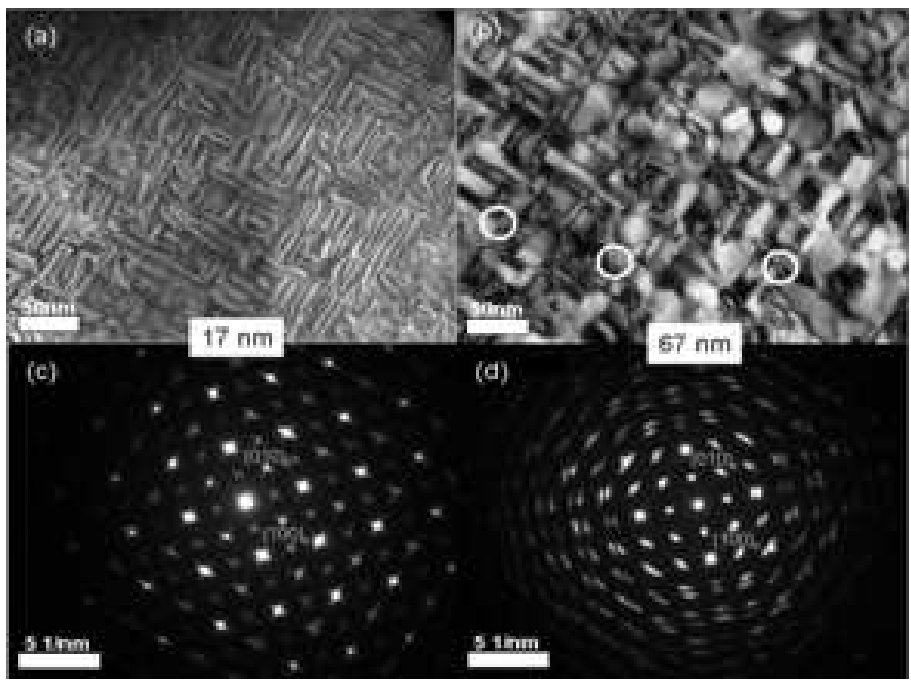}\\
\caption{Plan view TEM pictures of a 17nm (a) and 67nm (b)TMO film. Electron diffraction patterns of the same films are shown in
(c) and (d), respectively. The white circles shows show the presence of dislocations.}\label{Figure11_0001}
\end{figure}

\section{Concluding remarks}

We have successfully deposited epitaxial orthorhombic TbMnO$_3$ films on (001)-SrTiO$_3$, with thicknesses ranging from 8nm
to 80nm. The crystal structure of the thin strained films has been identified as a less distorted orthorhombic unit cell compared to the bulk one with a$_{film}$$>$a$_{bulk}$, b$_{film}$$<$b$_{bulk}$ and c$_{film}$$> $c$_{bulk}$. We found that the films grow with the c-axis out of plane and they orient themselves in the plane such that the $\langle$110$\rangle$$_o$ directions align with the $\langle$100$\rangle$$_c$ directions of the substrate. The orthorhombic lattice parameters ${a_o}$ and ${b_o}$ are constraint by epitaxy so that ${a_o}$$^2$+${b_o}$$^2$= (2a$_{STO}$)$^2$. This allows for four equivalent orientations and, therefore, four types of domains are observed, such that the in-plane diffraction patterns display four-fold symmetry.

With increasing thickness, the in-plane orthorhombic axes gradually change towards their bulk values, increasing the orthorhombicity of the films, but still keeping the partial coherence and epitaxy relation ${a_o}$$^2$+${b_o}$$^2$= (2a$_{STO}$)$^2$. Due to this, the amount of in-plane compression (and hence out-of-plane elongation) remains unchanged up to thicknesses of about 70nm. Above this thickness, the lattice relaxes to the bulk one. In spite of the unchanged in-plane compression, the thicker films do show a distribution of unit cell rotations coupled with increasing orthorhombic distortion.

It has been shown\cite{Rubi} that the physical properties of these films are very different from those of the bulk: The films show ferromagnetic interactions and spin-glass-like behavior below the N$\acute{e}$el temperature of $\sim 40K$, which are absent in the bulk. This can be of interest due to the scarcity of ferromagnetic insulators. Similar results have recently been obtained for this and other orthorhombic perovskites\cite{Marti,Kirby}. Here, we propose that this is directly associated to the domain walls, showing evidence that these constitute a large part of the volume of the films. On the other hand, the magnetic anomaly observed in the bulk material at about 28K, related to the transition to the spiral spin structure and the onset of ferroelectricity, could not be observed in these films\cite{Rubi}. This may be due to the decrease of in-plane anisotropy that we show takes place with decreasing thickness, or to the small domain size. Work is in progress to investigate these possibilities.

\section{Acknowledgements}
The authors would like to thank Umut Adem, Graeme Blake, Claire Colin, Maxim Mostovoy, Mufti Nandang, Gwilherm Nenert, Thom Palstra, Gijsbert Rispens, Oliver Seeck and Ard Vlooswijk for useful discussions and, very specially, Henk Bruinenberg, for his invaluable technical support. Finally, financial support by the EU STREP project MaCoMuFi(FP6-03321) is gratefully acknowledged.

\end{document}